\documentstyle[12pt]{article}
\begin{document}
\baselineskip=13pt
{\bf Possible Existence Of Topological Excitations In Quantum Spin Models In Low Dimensions}
\begin{center}
{\bf Ranjan Chaudhury}$^a$ and {\bf Samir K. Paul}$^b$\\
S. N. Bose National Centre For Basic Sciences\\
Block-JD, Sector-III, Salt Lake\\
Calcutta-700091,  India
\end{center}

\vspace{0.5cm}

\noindent        {\bf Abstract}

     The possibility of existence of topological excitations in the anisotropic quantum Heisenberg model in one and two spatial dimensions is studied using
 coherent state method. It is found that a part of the Wess-Zumino term
 contributes to the partition function, as a topological term for ferromagnets
 in the long wavelength limit in both one and two dimensions. In particular,
 the XY limit of the two-dimensional anisotropic ferromagnet is shown to retain
 the topological excitations, as expected from the quantum Kosterlitz-Thouless
 scenario.

\vspace{0.5cm}

a)    {\bf ranjan@boson.bose.res.in}

b)    {\bf smr@boson.bose.res.in}

\vspace{0.5cm}

\noindent        {\bf Introduction}

 Quantum spin systems in low dimensions have aquired considerable significance 
in condensed matter Physics in recent times.In particular in two dimensional  
(2D) spin $\frac{1}{2}$ quantum Heisenberg antiferromagnet (QHAF) evoked a lot
 of interest in the light of discovery of high-temperature superconductors$^1$.
 Many interesting theoretical and experimental work probing the magnetic
 property of various 2D systems and also that of many quasi-one dimensional
 (1D) systems brought into notice important features of anisotropic quantum
 spin models$^{2-4}$. Parallel to this, possible extention of
 Kosterlitz-Thouless(KT) scenerio to quantum ferromagnetic spin models has also been attempted$^5$.

        However, in spite of this endeavour many crucial questions have
 remained unanswered and in particular the origin of existence of topological
 excitations in quantum ferromagnetic and antiferromagnetic models seems to be
 mysterious   . The existence of topological excitations in isotropic 1D-AF is
 well known$^{3,6}$.The case of 1D ferro (both isotropic and anisotropic ) on
 the other hand has  drawn lesser attention$^{3,6}$.One possible reason for
this could be the lack of proper theoretical analysis of the quantum nature of 
the problem which we describe in this report. In the 2D case even for AF ,the
issue of existence
 of topological excitations is still not settled fully although most of the
 theoretical calculations rule out such a possibility$^{6,7}$.Moreover, the high
T$_c$ oxides in the insulating antiferromagnetic (AF) phase seem to be governed
 by anisotropic (2D) Heisenberg model, whereas the theoretical efforts have
 mostly been confined to the isotropic case only$^{1,2,6,7}$.the 2D-ferro
 situation remained even less understood so far$^{5,6}$.

        This motivated us to study the anisotropic quantum Heisenberg ferro and
antiferromagnetic models in 1D and 2D,in a unified manner.

\vspace{0.5cm}

\noindent       {\bf Mathematical Formulation}

\vspace{0.5cm}
  We analyse the quantum actions for XXZ ferromagnets and anti-ferromagnets in
1D and 2D by spin coherent state method$^6$.The philosophy behind this procedure
is that,the existence of topological term in the full quantum partition functionof a quantum spin system implies topological excitations in the system$^8$.
Keepingin mind the physically relevant situations , we choose the anisotropy of 
the above spin models to be XY-like.

     In the following we perform all the calculations on the lattice with a
finite lattice parameter 'a' in the long wavelength limit.We write down the
expressions of quantum Euclidean action in the quasicontinuum limit, so that we
have a clear understanding of the topological terms while the physical system
retains its lattice structure.

\vspace{0.5cm}

\noindent   {\bf Calculations}

\vspace{0.5cm}

        The quantum Euclidean action  $S_{E}[{\bf n}]$ for the spin coherent
fields ${\bf n} (t)$ can be written as$^{6,9}$,
\begin{equation}
S_{E} [{\bf n}] = -is S_{WZ}[{\bf n}] +\frac{s\delta t}{4} \int^\beta_0 dt 
\partial_t {{\bf n}(t)}^2 + \int^\beta_0 dt H({\bf n})
\end{equation}   
where s is the magnitude of the spin and
\begin{equation}
H({\bf n}) = \langle {\bf n} | H({\bf s}) |{\bf n}\rangle
\end{equation} 
$H({\bf s})$ being the spin Hamiltonian in the representation s.The Wess-Zumino
term  $S_{WZ}$ is given by$^6$
\begin{equation}
S_{WZ}[\bf n] = \int^\beta_0 dt \int^1_0 d\tau {\bf n}(t,\tau ) \cdot 
\partial_t {\bf n}(t,\tau ) \wedge \partial_\tau {\bf n}(t,\tau ) = A
\end{equation} 
with ${\bf n}(t,0) \equiv {\bf n}(t)$,${\bf n}(t,1) \equiv \bf n_0$,
${\bf n}(0, \tau ) \equiv {\bf n}(\beta ,\tau )$; $t\in [0,\beta ]$,
$\tau\in [0,1]$.

  In (3) A  is the area of the cap bounded by the trajectory   $\Gamma$
 parametrized by ${\bf n}(t)$ on the sphere

\begin{equation}
{\bf n} \cdot {\bf n} = 1
\end{equation}

Here  $|{\bf n}\rangle$   is the spin coherent state as defined in ref.(6).The
spin Hamiltonian for XXZ Heisenberg ferromagnet is given by

\begin{equation}
H({\bf S} ) = - g{\sum_{\langle {\bf r},{\bf {r\prime}} \rangle}} {\tilde
{\bf S}}({\bf r}) \cdot {\tilde{\bf S}}({\bf r}\prime) - g{\lambda_z}
{\sum_{\langle {\bf r},{\bf {r\prime}} \rangle}} {S_z}({\bf r}) {S_z}
({\bf {r\prime}})   
\end{equation}
with $g\ge 0$, $0\le \lambda_z \le 1$, where $\bf r$ and $\bf {r\prime}$ run
over the lattice and $\langle {\bf r},{\bf r}\prime \rangle$ signifies nearest
neighbour interaction and ${\bf S} = (\tilde{\bf S}, S_z)$.

\vspace{0.5cm}
\noindent  (i) {\bf Linear Chain}

    The quantum Euclidean action in the quasi-continuum limit can be written
as :
\begin{eqnarray}
{S_E}[\bf n] & = & -i s S^{NTOP}_{WZ} - {\frac{is}{2}}{\int^L_{-L}} dx 
{\int^{\beta}_0} dt \bf n \cdot {\partial_t}{\bf n}\wedge {\partial_x}{\bf n} 
+{\frac{s \delta t}{2a}} {\int^L_{-L}} dx {\int^\beta_0} dt\nonumber\\ 
&  &{({{\partial_t}\bf n})^2} + {\frac{{s^2}ga}{2}}{{\int^L_{-L}} dx}
{{\int^\beta_0} dt}[{({\partial_x}{\tilde{\bf n}})^2}\nonumber\\ 
&  & + {\lambda_z}({\partial_x}{n_z})^2]
\end{eqnarray}

where,${S^{NTOP}_{WZ}} = 2 {\sum_WZ}[{\bf n}(2ar)]$ is the nontopological part

of the WZ term$S_{WZ}$ on the chain, $L= 2ma$ and ${\bf n} = (\tilde{\bf n},n_z)$.
We have written down
eq.(6) in the long wavelength limit.We analyse $S_{WZ}$term in the following
manner :
\begin{eqnarray}
{\sum_{\bf r}} {S_WZ}[{\bf n}(2ar)]\nonumber\\ 
& = &{\sum_{r=-m}^m}{S_WZ}[{\bf n}(2ar)] + {\sum_{r=-m}^{m-1}} {S_WZ}
 [{\bf n}{{(2r+1)a}}]\nonumber\\
& = & 2 {S_WZ}[{\bf n}(-2am)] + {S_WZ}[{\bf n}(2am)] + 2 {\sum_{r=-m+1}^{m-1}} 
{S_WZ}[{\bf n}(2ar)]\nonumber\\ 
&   & + \frac{1}{2}{\int^L_{-L}} dx {\int^{\beta }_0} dt 
 {\bf n}\cdot {\partial_t}{\bf n} \wedge {\partial_x}{\bf n}\nonumber\\
& = & 2 {S_WZ}[{\bf n}(-2am)] + {S_WZ}[{\bf n}(2am)] + 2 {\sum_{r=-m}^{m-p}} 
{S_WZ}[{\bf n}(2ar)] + 2 [{\int_{-L}^{L-(p-1)2a}} dx\nonumber\\ 
&   & + {\int_{-L}^{L-(p-2)2a}} dx+ \cdots \nonumber\\
&   & + {\int_{-L}^{L-2a}} dx]{\int_0^{\beta }} dt {\bf n}\cdot {\partial_t}
{\bf n}\wedge {\partial_x}{\bf n}\nonumber\\ 
&   & + {\frac{1}{2}}{\int^L_{-L}} dx 
{\int^{\beta }_0} dt {\bf n}\cdot {\partial_t}{\bf n}\wedge {\partial_x}{\bf n}
\end{eqnarray}

Wher $1{\leq } p {\leq }2m$ . Since we keep the lattice parameter 'a' finite and we take $\delta t \rightarrow 0$ ,eq(6) takes the form :
\begin{equation}
{S_E}[{\bf n}] = - i s {S_{WZ}^{NTOP}} - {\frac{is}{2}}{\int^L_{-L}} dx 
{\int^{\beta }_0} dt {\bf n}\cdot {\partial_t}{\bf n}\wedge {\partial_x}{\bf n}
+ {\frac{{s^2}ga}{2}}{\int^L_{-L}} dx {\int^{\beta }_0} dt [{({\partial_x}
{\tilde{\bf n}})^2} + {\lambda_z}{({\partial_x}{n_z})^2}]
\end{equation}
In order that the Euclidean action in (8) is finite for very large L we have :
\begin{equation}
{\lim_{{\left|{\bf x}\right|}\to\infty }}{\partial_x}{\tilde{\bf n}} = 0 ,
{\lim_{{\left|{\bf x}\right|}\to\infty }}{\partial_x}{n_z} = 0
i.e.{\bf n}\longrightarrow {{\bf n}_0}(t) on the circle 
{x^2} + {t^2} = {R^2};R\rightarrow \infty with
{\bf n}\cdot {\bf n} = 1
\end{equation}
 This defines a mapping from (x,t)-space to the internal space ${\bf n}\cdot 
 {\bf n} = 1$.However, there is a smaller class of $\bf n$ fields on the
(x,t)-space which satisfies (9) with ${\bf n}_0 (t)$ independent of t, denoted
by ${\bf n}_0$. In that case the  boundary points in the (x,t)-space are
identified to a single point and we have a topological mapping
${S^2_{Phys.}}\longrightarrow {S^2_{Int.}}$ with ${\Pi_2}({S^2}) = Z^{10}$.
The winding number in this case is given by$^8$
\begin{equation}
Q = {\frac{1}{4\pi }} \int dx dt {\bf n}\cdot {\partial_t}{\bf n} \wedge 
{\partial_x}{\bf n}
\end{equation}
where $Q\in z$.Thus for field configurations represented by
\begin{equation}
\{ {\bf n}(x,t) : {\lim_{{\left|{\bf x}\right|}\to\infty}} {\bf n}(x,t)
= {{\bf n}_0}\}
\end{equation}
eqn(7) can be written as:
\begin{eqnarray}
{\sum_{\bf r}} {S_{WZ}} [{\bf n}(2ar)]
& = & 2 {\sum^{m-p}_{r=-m}} {S_{WZ}} [{\bf n}(2ar)]
+2 [{\int^{L-(p-1)2a}_{-L}} dx + {\int^{L-(p-2)2a}_{-L}} dx+ \cdots\nonumber\\ 
&   & + {\int^{L-2a}_L}]
{\int^\beta_0} dt {\bf n}\cdot {\partial_t}{\bf n}\wedge {\partial_x}
{\bf n}\nonumber\\ 
&   & + {\frac{1}{2}}{\int^L_{-L}} dx {\int^\beta_0} dt {\bf n}\cdot 
{\partial_t}{\bf n}\wedge {\partial_x}{\bf n}
\end{eqnarray}
where we have made use of eq.(3).Notice that only the last term in eq.(12)
covers the entire chain where the boundary points are identified via the 
configuration (11).Therefore the topological content in $S_{WZ}$ is the last
integral in (12) which is same as (10).The rest of the term in (12) are
non-topological.From eq.(7) the non-topological part can be written as:
\begin{equation}
{S^{NTOP}_{WZ}} = 2 {\sum^{m-1}_{r=-m}} {S_{WZ}} [{\bf n}(2ar)]
\end{equation}
Due to translational invariance of $\bf n$ on the chain each term in (13)
describes the same cap of area A given by eq.(3).Thus (13) can be written as:
\begin{equation}
{S^{NTOP}_{WZ}} = 2 {\sum^{m-1}_{r=-m}} A = 2 (2m-1) A = constant
\end{equation}
Therefore eq.(8) becomes:
\begin{equation}
{S_E}[{\bf n}] = - {\frac{i s}{2}}{\int^L_{-L}} dx {\int^\beta_0} dt 
{\bf n}\cdot {\partial_t}{\bf n}\wedge {\partial_x}{\bf n}
+ {\frac{{s^2}ga}{2}} {\int^L_{-L}} dx {\int^\beta_0} dt
[{({\partial_x}{\tilde{\bf n}})^2} + {\lambda_z} {({\partial_x}{n_z})^2}]
\end{equation}
The eq.(12) corresponding to an antiferromagnet ($g\le 0$) reads
\begin{equation}
{S_{WZ}} = - {\frac{1}{2}} {\int^L_{-L}} dx {\int^\beta_0} dt 
{\bf n}\cdot {\partial_t}{\bf n}\wedge {\partial_x}{\bf n}
\end{equation}
with $S^{NTOP}_{WZ}$ vanishing due to staggering operation.So we get back the
same result as obtained for isotropic antiferromagnet$^6$.

\vspace{0.5cm}
\noindent  (ii) {\bf 2D Square Lattice}

       The spin Hamiltonian in this case is given by (5) where ${\bf r}$ runs 
over the 2D-square lattice.The quantum action for the anisotropic ferromagnet in the
long wavelength limit is given by:
\begin{equation}
{S_E}[{\bf n}] = -i s {\sum_{\bf n}} {S_{WZ}} [{\bf n}({\bf r})] + 
+ {\frac{g{s^2}}{2}} {\int^L_{-L}} dx dy {\int^\beta_0} dt 
[{({\partial_x}{\tilde{\bf n}})^2} + {\lambda_z}{({\partial_x}{n_z})^2} + 
{({\partial_y}{\tilde{\bf n}})^2} + {\lambda_z}{({\partial_y}{n^z})^2}](x,y,z)
\end{equation}
Finiteness of the action (17) gives :
\begin{equation}
{\bf n}(x,y,t)\rightarrow {{\bf n}_0}(t) on  the  2-sphere   {{x^2}+{y^2}+{z^2}
= {R^2}} , R\rightarrow \infty and {\bf n}\cdot {\bf n} = 1
\end{equation}
This boundary condition gives a mapping from the (x,y,t)-space to the internal
sphere ${\bf n} \cdot {\bf n} = 1$. However, (18) admits a class of $\bf n$
fields where ${\bf n}_0 (t)$ is independent of t denoted by  ${\bf n}_0$ in
which case we have a mapping ${S^3_{Phys}}\longrightarrow {S^2_{Int}}$ and
${\Pi_3}({S^2}) = Z^{10}$. Thus for the field configurations
\begin{equation}
\{ {\bf n}(x,y,t) : \lim_{{\left|x\right|}\to\infty}
{\bf n}(x,y,t) = {\bf n}_0\}
\end{equation}
Following the same line of arguments as  linear chain we can show that for
field configurations satisfying (19) the $S_{WZ}$ term corresponding to the
2D square lattice can be written as :
\begin{equation}
{\sum_{\bf r}} {S_{WZ}}[{\bf n}({\bf r})]
= {S_{WZ}^{NTOP}} + \frac{1}{2a} {\int^L_{-L}} dx dy {\int^\beta_0} dt 
[{\bf n}\cdot {\partial_t}{\bf n}\wedge {\partial_x}{\bf n}
+ {\bf n}\cdot {\partial_t}{\bf n}\wedge {\partial_y}{\bf n}](x,y,t)
\end{equation}
\begin{equation}
 {S_{WZ}^{NTOP}} = {\{2 (2m-1)\}^2} A
\end{equation}
For the mapping ${S^3_{Phys}}\longrightarrow {S^2_{int}}$, we can parametrize
the (x,y,t)-space with boundary points identified by y-planes in which case
$\frac{1}{4\Pi}{\int^L_{-L}} dx {\int^{\beta}_0} dt {\bf n}\cdot 
{\partial_t}{\bf n}\wedge {\partial_x}{\bf n}$ will be a winding number through
eq(10) for each y.This happens because of the fact that each (x,t)-plane
(i.e $y = constant $) has its boundary points identified for field 
configurations satisfying (19) and the (x,t)-plane can be thought as a sphere
$S^2$ passing through the north pole of $S^3$. So for each y we have a mapping
from the corresponding (x,t)-plane ($=S^2$) to $S^2_{Int}$ .Therefore we can
 write
\begin{equation}
\frac{1}{4\pi}{\int^L_{-L}} dx dy {\int^\beta_0} {\bf n}\cdot
{\partial_t}{\bf n}\wedge {\partial_x}{\bf n}(x,y,t) = {\int^L_{-L}} dy Q(y)
\end{equation}
By similar arguments :
\begin{equation}
\frac{1}{4\pi}{\int^L_{-L}} dx dy {\int^\beta_0} dt {\bf n}\cdot 
 {\partial_t}{\bf n}\wedge {\partial_y}{\bf n}(x,y,t) = {\int^L_{-L}} dx Q(x)
\end{equation}
Note that we can apply the above principle for the mapping
${S^3_{Phys}}\longrightarrow {S^2_{Int}}$ since ${{\Pi}_3}({S^2}) = Z^{10}$. 

  Thus using eq(20),(21) the action (17) becomes :
\begin{eqnarray}
{S_E}[{\bf n}] & = & - \frac{is}{2a} {\int^L_{-L}} dx dy {\int^\beta_0} dt 
[{\bf n}\cdot {\partial_t}{\bf n}\wedge {\partial_x}{\bf n} +\nonumber\\ 
&   & {\bf n}\cdot {\partial_t}{\bf n}\wedge {\partial_y}{\bf n}](x,y,t) + 
{\frac{g{s^2}}{2}}{\int^L_{-L}} dx dy {\int^\beta_0} dt\nonumber\\ 
&   & [ {({\partial_x}{\tilde{\bf n}})^2} + {\lambda_z}{({\partial_x}{n_z})^2} 
+ {({\partial_y}{\tilde{\bf n}})^2} + {\lambda_z}{({\partial_y}{n_z})^2}]
(x,y,t)
\end{eqnarray}
The first integral in (24) is a topological term as follows from (22) and (23).
Let us point out that the right hand side of eq(20) vanishes identically on the
long wave length limit under staggering operation in the case of
 antiferromagnets .

\vspace{0.5cm}

\noindent       {\bf Conclusions}

           We have presented a unified scheme for analysing the topological
terms in the effective action corresponding to the long wavelength limit
of XY-like anisotropic quantum Heisenberg ferro and antiferro-magnets in one
and two spatial dimensions for any value of the spin. Our calculation 
brings out clearly the hidden topological contribution from $S_{WZ}$ term, which
influences the statistical mechanics of ferromagnets in one dimension. This
is probably manifested in the 'soliton-like excitations' occuring in many
experimental systems corresponding to these models$^{3,4}$. It may also be 
interesting to point out that in 1D the roles of kink and anti-kink are inter-
changed as we go from ferro to anti-ferro due to sign reversal in the
respective topological terms [see eqs. (15) and (16)]. In 2D situation in
the limit ${\lambda_z}\rightarrow 0$, these excitations probably lead to the
proposed "vortex-antivortex" scenario in the 'quantum KT' picture$^5$. On the
contrary the 2D-AF model does not exhibit any topological excitation in its
long wavelength behaviour.
            Let us conclude by pointing out that our whole calculational
approach is meaningful only in the low temperature regime where the spin-spin
correlation length is appreciably large$^{6,11}$.
\vspace{0.5cm}

\noindent          {\bf References}
\vspace{0.5cm}

\noindent 1

 P.W. Anderson, Science, 235, 1196 (1987) ; P.W. Anderson, Phys. Rev. Lett.,
59, 2407 (1987) ; J.G. Bednorz and K.A. Muller, Z. Phys., B64, 189 (1986) ;
P. Chu et al, Phys. Rev. Lett., 58, 405 (1987).

\vspace{0.5cm}

\noindent 2

 Y. Endoh et al, Phys. Rev., B 37, 7443 (1988) ; K. Yamada et al, Phys. Rev.,
B 40, 4557 (1989) ; M. Sato et al, Phys. Rev. Lett., 61, 1317 (1988) ; 
R. Chaudhury, Ind. J. Phys. (Special issue on High Temperature Superconduc-    
tivity) 66A, 159 (1992).

\vspace{0.5cm}

\noindent 3

 G.M. Wysin and A.R. Bishop, Phys. Rev., B 34, 3377 (1986) ; H.J. Mikeska,
J. Phys. C, 13, 2913 (1980) ; J.des Cloizeaux and J.J. Pearson, Phys. Rev.,
128, 2131 (1967) ; Y. Endoh et al, Phys. Rev. Lett., 32, 170 (1974) ;
M. Imada, 'Finite Temperature Excitations of the XYZ Spin Chain' (ISSP,
Tokyo, 1982).

\vspace{0.5cm}

\noindent 4

 K. Hirakawa, H. Yoshizawa and K. Ubukoshi, J. Phys. Soc. Jpn., 51, 2151
(1982) ; K. Hirakawa et al, J. Phys. Soc. Jpn., 52, 4220 (1983) ;
S. Komineas and N. Papanicolaou (Private Communications).

\vspace{0.5cm}

\noindent 5

 J.M. Kosterlitz and D.J. Thouless, J. Phys. C, 6, 1181 (1973) ; F.G.
Mertens et al Phys. Rev. Lett., 59, 117 (1987) ; E. Loh, Jr., D.J.
Scalapino and P.M. Grant, Phys. Rev., B31, 4712 (1985) ; F. Fucito and 
S. Solomon, DOE Research and Development Report, CALT-68-1023 (1981).

\vspace{0.5cm}

\noindent 6
 
 E. Fradkin and M. Stone, Phys. Rev., B 38, 7215 (1988) ; E. Fradkin,
'Field Theories of Condensed Matter Systems' (Addison Wesley, California,
1991).

\vspace{0.5cm}

\noindent 7

 P. Horsch in Lecture Note of the Mini-Workshop on ' Mechanisms for High
Temperature Superconductivity ' (ICTP, Trieste, 1988) ; S. Chakraverty,
B.I. Halperin and D.R. Nelson, Phys. Rev. Lett., 60, 1057 (1988) ; S. Tyc,
B.I. Halperin and S. Chakraverty, 62, 835 (1989) ; A. Auerbach and D.P.
Arovas, Phys. Rev. Lett., 61, 617 (1988).

\vspace{0.5cm}

\noindent 8

 R. Rajaraman, 'Solitons and Instantons : An introduction to solitons and
instantons in quantum field theory ' (North-Holland, Amsterdam, 1982).

\vspace{0.5cm}

\noindent 9

 A. Parola, Private Communications (1988).

\vspace{0.5cm}

\noindent 10

 N. Steenrod, 'The Topology of Fibre Bundles ' (Princeton University Press,
New Jersey, 1951).

\vspace{0.5cm}

\noindent 11

 C.K. Majumdar in Proceedings of the Winter School and International
Colloquium held at Panchgani, India, edited by B.S. Shastry, S.S. Jha and
V. Singh (Springer-Verlag, Berlin Heidelberg, 1985), 142.

\end{document}